\documentclass[twocolumn,amsmath,amssymb,floatfix,prb,epsf,showpacs]{revtex4}
\usepackage{amsmath,amssymb,natbib,bm,graphicx,url,epsfig}
\usepackage[ansinew]{inputenc}

\newcommand{\be}{\begin{equation}}
\newcommand{\ee}{\end{equation}}
\newcommand{\bes}{\begin{equation}\begin{split}}
\newcommand{\ees}{\end{split}\end{equation}}

\begin{document}
\title{Penetration of external field into regular and random arrays of
nanotubes: Implications for field emission}

\author{T. A. Sedrakyan, E. G. Mishchenko, and M. E. Raikh}

\address{Department of Physics, University of Utah, Salt Lake
City, UT 84112}

\date{\today}

\begin{abstract}
We develop an analytical theory of polarization of a vertically
aligned  array of carbon nanotubes (NTs) in external electric
field. Such arrays are commonly  utilized in field-emission
devices,  due to the known electrostatic effect of strong field
enhancement near the tip of an {\em individual} NT. Small ratio of
the NT radius to the separation between neighboring NTs allows to
obtain asymptotically exact solution for the distribution of
induced charge density along the NT axes. For a regular array,
this solution allows to trace the suppression of the field
penetration with increasing the density of NTs in the array. We
demonstrate that for a random array, fluctuations in the NT
density terminate the applicability of our result at distances
from the NT tips much larger than the field penetration depth,
where the induced charge density is already exponentially small.
Our prime conclusion is that, due to {\em collective} screening of
the external field by the array, the field-emission current {\em
decreases} drastically for dense arrays compared to an individual
NT. We argue that the reason why the strong field emission,
described by the Fowler-Nordheim law and observed in realistic
arrays, is the strong dispersion in heights of the constituting
NTs.
\end{abstract}


\maketitle

\section{Introduction}

First report of field emission from carbon NTs had appeared a
decade ago\cite{Science0}.  It was followed by a demonstration
\cite{sourse} that  arrays of NTs can be patterned into emitting
and non-emitting regions. Since then, the field emission
properties of carbon NTs command a steady interest from
researchers worldwide. The uniqueness of these properties
originates from  geometry of a NT. Namely, due to a small NT
radius, $r$, the electric field applied between the substrate
(cathode), on which the NTs are grown, and the anode is enhanced
by a large factor $\beta \gg 1$ near the nanotube tip. Such an
enhancement translates into high probability of electron
tunnelling toward the anode, leading to desirable low turn-on
voltage for field emission. This property, combined with high
emission current density, made possible a successful fabrication
of the row-column matrix-addressable flat panel display based on
carbon NTs
\cite{flat0,flat1,flat2,flat3,flat4,flat5,flat6,flat7,flat8}.
Currently, flat panel displays constitute one of the most
prominent applications of NTs \cite{baughman02}. Geometrical
characteristics of individual NTs utilized in the first display
\cite{flat0} were highly dispersed. Further advances in
fabrication \cite{Science1} allowed to achieve excellent vertical
alignment and high homogeneity in the lengths and radii of NTs
\cite{latest1,latest3,latest2,latest4}. This suggests that  NTs on
the cathode of a
 display can, in the first approximation, be viewed as constituting
a {\em regular array} of identical NTs. Such an array is
schematically illustrated in Fig.~1.

%%%%%%%%%%%%%%%%%%%%%%%%%%%%%%%%%%%%%%%%%%%%%%%%%%%%%%%%%%%%%%%%%%%%%%%%%
%%%%%%%%%%%%%%%%%%%%%%%%%%%%%%%%%%%%%%%%%%%%%%%%%%%%%%%%%%%%%%%%%%%%%%%%

\begin{figure}[t]
\centerline{\includegraphics[width=90mm,angle=0,clip]{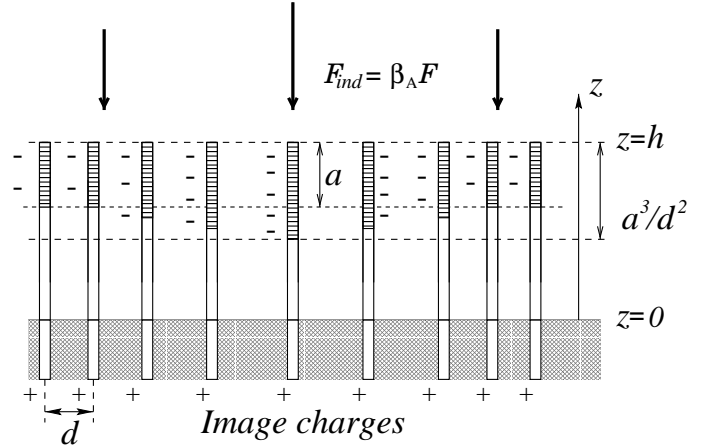}}
\caption{Schematic illustration of the forrest of NTs of a height,
$h$, grown on the substrate, $z=0$. The average distance between
neighboring  NTs is $d$. Shaded regions at the NT tips illustrate
the charges, induced in the NTs by the external field, $F$. For a
dense forrest, the typical penetration depth, $a$, exceeds $d$.
Vertical arrows illustrate  electric field, $F_{ind}$, created by
induced charges. Regions of lower NT density correspond to higher
field-enhancement-factor, $\beta_A=F_{ind}/F$. In these regions,
external field penetrates deeper into the forrest. Fluctuations in
the induced charge density, due to the randomness in the NT
positions, exceed the average density at depth $(h-z)>a^3/d^2$.}
\label{fig:1}
\end{figure}

%%%%%%%%%%%%%%%%%%%%%%%%%%%%%%%%%%%%%%%%%%%%%%%%%%%%%%%%%%%%%%%%%%%%%%%%%
%%%%%%%%%%%%%%%%%%%%%%%%%%%%%%%%%%%%%%%%%%%%%%%%%%%%%%%%%%%%%%%%%%%%%%%%%%
%%%%%%%%%%%%%%%%%%%%%%%%%%%%%%%%%%%%%%%%%%%%%%%%%%%%%%%%%%%%%%%%%%%%%%%%%%

On the theoretical side, the focus of the previous studies
\cite{theory1,theory2,theory3,theory4,theory5,theory6,theory7,theory8,theory9,theory10}
of field emission from nanotubes was the effect of band structure
and tip geometry of an {\em individual} NT on the emission
probability. These studies left out the fact that  {\em all} NTs
are coupled to each other {\em electrostatically}.

The main point of the present paper is that, for a regular NT
array mutual electrostatic  coupling  of NTs in the array has a
dramatic effect on the field emission, especially in dense arrays.
By dense we mean the arrays in which the separation, $d$, between
neighboring NTs is much smaller than the NT height, $h$. The
situation $d\ll h$ is quite common in realistic field-emission
setups, see, {\em e.g.,}
Refs.~\onlinecite{latest3,latest2,latest4}.
%With a typical size of a display  pixel $\sim 0.1$ mm and minimal
%areal
%density of nanotubes $\sim 10^8$ cm$^{-2}$, each pixel
%contains large number of NT.
To substantiate this point, consider first two parallel NTs
separated by $d\ll h$ in the external electric field, $F$,
directed along their axes, see Fig.~1. It might seem, that, if $d$
exceeds the tunnelling length for field emission from each of the
NTs, then both NTs emit independently. This, however, is not the
case. The reason is that the enhancement of the electric field
near the tip of each NT is governed by the charge density, induced
by the external field. For two parallel NTs, the induced charge
density {\em per} NT is approximately {\em two times smaller} than
for an isolated NT.\cite{remark} As a result, the field
enhancement near the tip of each NT becomes  smaller due to the
presence of the neighbor.

Compared to a pair of neighboring NTs, the suppression of the
field enhancement becomes much more pronounced {\em in the NT
array}.  On the qualitative level, this conclusion follows from
the fact that each NT in the array interacts with $\sim (h/d)^2
\gg 1$ neighbors. As we will demonstrate below, external field
simply {\em does not penetrate} into the sufficiently dense array.
On the intuitive level, this trend has been previously
understood.\cite{Nilsson, latest2, Manohara} In particular, in
Ref.~\onlinecite{Nilsson} numerical simulations illustrating the
suppression of the field enhancement for three parallel NTs with
decreasing distance between them were reported.  However, full
understanding of this suppression {\em in the array} with
arbitrary ratio $d/h$ requires
%development of
an analytical theory. Such a theory is developed in the present
paper. In particular we demonstrate that

{\em i)} penetration of the external field into the array is
described by a simple  function, $\sinh(z/a)$, where the
penetration depth, $a$, is much smaller than $h$ for a dense
array, but still much bigger than $d$;

{\em ii)} distributions of induced charge density in a regular and
 completely random NT arrays are approximately {\em the same};

{\em iii)} with regard to the field emission, the enhancement of
external field for the array, as compared to the individual NT, is
suppressed by the factor $\approx (h/a)$.

The reason why the electrostatic problem, in which the variables
cannot be separated, allows an asymptotic analytical solution is
the presence of large parameters $(h/r) \gg 1$ and $(d/r) \gg 1$.
As it was demonstrated in Ref. \onlinecite{we}, for a single NT,
the relation $h\gg r$ allows one to find analytically the
distribution of induced charges in external field. Here the
approach of Ref. \onlinecite{we} is generalized to the NT array.

The paper is organized as follows. In Sect.~II we review the
Thomas-Fermi description \cite{we} of polarization of a single NT
in external field. In Sect.~III  we generalize the Thomas-Fermi
equation to the NT array. In Sections IV-V we analyze this
equation for a regular array, and find its asymptotic [in the
parameter $(d/r)\gg 1$] solution. Robustness of this solution with
respect to randomness in the NT positions is demonstrated in
Sect.~VI. In Sect.~VII we apply to the obtained solution for
distribution of the induced charge density to calculate the field
emission current from the array. Relation of our theory to
experiment is addressed in Sect.~VIII.

\section{Single NT}
Denote with $\rho(z)$ the {\em linear} charge density on the NT
surface at a distance $z<h$ from the substrate. The Thomas-Fermi
equation for $\rho(z)$ reads\cite{we}
%Apparently,
%$\rho(z)$ must exhibit itself as a solution of the following
%integral equation
\begin{eqnarray}
\label{basic} eFz=\frac{1}{g}~\rho (z)+\frac{1}{e}\int
_0^h\!dz^{\prime}\;{\cal S}_0(z,z^{\prime})\rho (z^{\prime}),
\end{eqnarray}
where the kernel, ${\cal S}_0(z,z^{\prime})$, is defined as
\begin{eqnarray}
\label{S} {\cal S}_0(z,z^{\prime})=\Phi (z-z^{\prime})
-\Phi(z+z^{\prime}).
\end{eqnarray}
Here the function $\Phi (z)$ is the interaction potential between
two points on the NT surface, separated vertically by $z$. It
represents an azimuthal average of the Coulomb potential
\begin{equation}
\label{interaction} \Phi(x)= \frac{e}{\pi}
\int_0^{\pi}\!\frac{d\alpha}
{\left[x^2+4r^2\sin^2(\alpha/2)\right]^{1/2}}~~,
\end{equation}
where $r$ is the NT radius. The second term in  Eq.~(\ref{S})
accounts for the image charges. The meaning of Eq.~(\ref{basic})
is the following. The lhs is the bare potential. The potential,
acting on  a given electron at the NT surface represents the sum
of this potential and of the potential, created by the induced
charges. This resulting potential defines the {\em local} value of
the Fermi energy, which, in turn, fixes the local value of the
Fermi momentum. This Fermi momentum, on the other hand, is
linearly proportional to the charge density in one dimension. This
is a standard reasoning behind the Thomas-Fermi description.
Within this description Eq.~(\ref{basic}) is nothing but the
condition, that the electrochemical potential remains constant
along the NT.

The prime simplification that allows analytical solution of
Eq.~(\ref{basic}) is that, with logarithmic accuracy, $\rho
(z^{\prime})$ in the integrand in the rhs of Eq.~(\ref{basic}) can
be replaced by $\rho (z)$. Upon this replacement, we have
\begin{eqnarray}
\label{approximate} \frac{1}{e}\int_0^h \!dz^{\prime}\;{\cal
S}_0(z,z^{\prime})\approx %\ln{\Biggl(\frac{h^2}{r^2}\Biggr)}
\ln{\frac{h^2}{r^2}} -
%\ln{\Biggl(\frac{h^2}{z^2}\Biggr)}
\ln{\frac{h^2}{z^2}} +\ln{\Biggl(\frac{1-z/h}{1+z/h}\Biggr)},
\end{eqnarray}
where we assumed that $(h-z) \gg r$. With the same logarithmic
accuracy, for $z\gg r$, the rhs of Eq.~(\ref{approximate}) can be
replaced by $2{\cal L}_h$, where ${\cal L}_h~=~\ln(h/r)$. Then we
obtain the following analytical solution for the induced charge
density\cite{we}
\begin{equation}
\label{single} \rho(z)\approx \frac{gFz}{1+2g{\cal L}_h},
\end{equation}
where we have introduced a dimensionless interaction parameter
$g=2Ne^2/\pi\hbar v_0$. The above result for $\rho(z)$ is
approximate, in the sense, that the numerical factor in the
argument of a large logarithm, ${\cal L}_h \gg 1$, is not
specified. Equation (\ref{approximate}) represents the solution of
Eq.~(\ref{basic}), which satisfies the obvious condition $\rho
(0)=0$. An improved analytical description yielding the result
coinciding with Eq.~(\ref{single}) in the limit of large ${\cal
L}_h$ was recently reported in Ref. \onlinecite{chinese}.

The remarkable property of the solution (\ref{single}) is that the
NT with poor ``metallicity'', $g<1$, eventually becomes metallic as
the length, $h$, of NT is increased. This is not the case if the
NT is located parallel to the conducting gate at distance $D\ll
h$. In the latter case,\cite{rotkin,blanter} one has to replace
${\cal L}_h$ by $\ln (D/r)$.

\section{Thomas-Fermi equation for an array}

Consider an array of parallel NTs  located at points, ${\bf R}_i$,
on the substrate, see Fig.~1.  To set the Thomas-Fermi equation
for a given NT, $i$, in the array, one has to take into account
that the external potential, leading to the charge separation,
contains, in addition to $eFz$, the potentials created by charges
induced on all other NTs. Then the generalized Eq.~(\ref{basic})
reads
\begin{eqnarray}
\label{generalized} eFz=\frac{1}{g}~\rho_i(z)+\int_0^{h}
dz^{\prime}\sum_{j}\rho_j(z^{\prime}) {\cal S}(z,z^{\prime};{\bf
R}_i-{\bf R}_j),
\end{eqnarray}
where the kernel, ${\cal S}$, is given by
\begin{eqnarray}
{\cal S}(z,z^{\prime};{\bf
R})&=&\frac{1}{\sqrt{(z-z^{\prime})^2+\vert{\bf
R}\vert^2}}\nonumber\\
&-& \frac{1}{\sqrt{(z+z^{\prime})^2+\vert{\bf R}\vert^2}}.\quad
\end{eqnarray}
It is  convenient to rewrite Eq.~(\ref{generalized}) in the
``continuous'' form by introducing the position-dependent density,
$\rho (z,{\bf R})$, and the local concentration of NTs
\begin{eqnarray}
\label{nn} {\cal N}({\bf R})=\sum_i\delta\left({\bf R}-{\bf
R}_i\right).
\end{eqnarray}
In the  new notations Eq.~(\ref{generalized}) takes the form
\begin{eqnarray}
\label{EQN1} eFz=\frac{1}{g}~\rho (z,{\bf R}) \qquad \qquad \qquad
\qquad \qquad
\qquad \nonumber \\
+\int d{\bf R}^{\prime} {\cal N}({\bf R^{\prime}})\!\!\int
_0^h\!\! dz^{\prime}\;{\cal S}(z,z^{\prime};{\bf R}-{\bf
R}^{\prime})\rho (z^{\prime},{\bf R}^{\prime}).\quad
\end{eqnarray}
The Thomas-Fermi equation in the form (\ref{EQN1}) is convenient
for further analysis. This is because, as we will see below, the
distributions, $\rho_i(z)$, are almost the same for {\em all} $i$
even for completely random array.
\section{Regular array}

Consider a regular array in the form of a square lattice with a
lattice constant, $d$. Then the coordinates of  ${\bf R}_i$ in
Eq.~(\ref{nn}) are $\{nd,md\}$ with integer $m$ and $n$.
Obviously, for the regular array the induced charge density,
$\rho(z)$, is the same for all NTs, so that Eq.~(\ref{EQN1})
acquires the form
 \begin{eqnarray}
\label{EQN11} eFz=\left(\frac{1+ 2g{\cal L}_{d}}{g} \right)~\rho
(z)+ \int _0^h\!\! dz^{\prime}\;{\cal S}_{ext}(z,z^{\prime})\rho
(z^{\prime}),\quad
\end{eqnarray}
where
\begin{eqnarray}
\label{array} {\cal
S}_{ext}(z,z^{\prime})=\sum_{\{m,n\}\neq\{0;0\}}\left[
\frac{1}{\sqrt{\left(z-z^{\prime}\right)^2+\left(m^2+n^2\right)d^2}}
\right. \nonumber \\
\left.
-\frac{1}{\sqrt{\left(z+z^{\prime}\right)^2+\left(m^2+n^2\right)d^2}}\right
].\qquad
\end{eqnarray}
Here we have isolated the self-action, $m=n=0$, of a  NT. For  a
single NT this self-action is described by a large logarithm
${\cal L}_h$. In the array, however, the interaction
 is screened at distances $\vert
z-z^{\prime}\vert \gtrsim d$ by the neighboring NTs. To account
for this screening, the logarithm,
 ${\cal L}_h$, in Eq.~(\ref{EQN11}) is replaced by ${\cal L}_d=\ln(d/r)$ .

It is apparent that both terms in the sum (\ref{array}) diverge
due to the contributions from large $m$, $n$. However, the
divergent parts in both terms cancel each other. Physically, this
reflects the screening by the image charges (see Fig. 1).
Replacing the sums over $m$ and $n$ in Eq.~(\ref{array}) by
integrals, we obtain the following expression for the kernel
${\cal S}_{ext}(z,z^{\prime})$
\begin{eqnarray}
\label{S-result} {\cal S}_{ext}(z,z^{\prime})= 2\pi {\cal
N}_0\Bigl[z+z^{\prime}-\vert z-z^{\prime}\vert\Bigr]=\nonumber\\
4\pi {\cal
N}_0\Bigl[z^{\prime}\Theta(z-z^{\prime})+z\Theta(z^{\prime}-z)\Bigr],
\end{eqnarray}
where $\Theta(x)$ is the step-function. Note that both steps,
replacing ${\cal L}_h$ in Eq.~(\ref{EQN11}) by ${\cal L}_d$ and
replacing the sums in Eq.~(\ref{array}) by integrals are by no
means obvious and require justification. This justification is
provided in the next Section. In the present Section we
demonstrate that the integral equation (\ref{EQN11}) with the
kernel (\ref{S-result}) can be solved analytically. Upon taking
the derivative from both sides of Eq.~(\ref{EQN11}), we obtain
\begin{eqnarray}
\label{equation} eF=\left(\frac{1+ 2g{\cal L}_{d}}{g}
\right)\frac{d\rho}{dz}+{4\pi}{\cal N}_0
\int_z^{h}\!\!dz^{\prime}\rho(z^{\prime}).
\end{eqnarray}
It is now easy to see that, within a factor, the first term in
Eq.~(\ref{equation}) is the second derivative of the second term.
Thus, Eq.~(\ref{equation}) can be viewed as a second-order
differential equation with respect to $\int_z^h\!\!
dz^{\prime}\rho(z^{\prime})$. The solution of this equation,
satisfying the condition $\rho(0)=0$, has the form
\begin{eqnarray}
\label{final1} \rho(z)= \rho_0\sinh \left(z/a\right),
\end{eqnarray}
where $\rho_0$ is defined as
\begin{eqnarray}
\label{final2} \rho_0=\frac{eFga}{\left(1+2g{\cal
L}_d\right)\cosh(h/a)},
\end{eqnarray}
and the length, $a$, is given by
\begin{eqnarray}
\label{depth} a=\frac{1}{2}\;\sqrt{\frac{1+2g{\cal L}_d}{\pi g
{\cal N}_0}}.
\end{eqnarray}
The above expression for the induced charge density constitutes
the main result of the present paper. It is seen from
Eq.~(\ref{final1}) that $a$ plays the role of the penetration
depth of the external field into the array. In the limit of very
low density,  ${\cal N}_0~=~d^{-2} \ll 1/h^2$,~~
Eqs.~(\ref{final1})--(\ref{depth}) reproduce the result
Eq.~(\ref{single}) for a single NT, as it could be expected, since
the mutual influence of  NTs, separated by a  distance  $\gtrsim
h$ is negligible.

 It also follows from Eqs.~(\ref{final1})--(\ref{depth}) that in the  limit of large  ${\cal
N}_0$, such that $a \ll h$, the induced charge density is
concentrated near the NT's tips and falls off towards the
substrate exponentially, as $\exp\{-(h-z)/a\}$. This weak
penetration of external field into the array
 is a consequence of the collective screening. Indeed,  in terms of
screening properties, the array of a high density can be viewed as
a homogeneous metallic plate. Our crucial observation, however, is
that the penetration depth exceeds {\em parametrically}  the
lattice constant, $d$, both for large and small values of the
interaction parameter, $g$.

By virtue of the relation $a/d >1$, there are {\em many} NTs
within the penetrations depth. This, in turn, suggests that
Eqs.~(\ref{final1})--(\ref{depth}) apply to the random array with
average areal concentration of NTs ${\cal N}_0=d^{-2}$.

\section{Derivation}

In the previous section the derivation  was based on two intuitive
assumptions. Namely, we have assumed the self-action of a NT in
the array is described by ${\cal L}_d$ instead of  ${\cal L}_h$
for an isolated NT, and that the sum over $m$ and $n$ in
Eq.~(\ref{array}) can be replaced by the integral. To justify
these assumptions, below we calculate the sum Eq.~(\ref{array})
more accurately.
%The prime step is to use the following (obvious) identity.
In order to do so, we employ the following (obvious) identity.
Consider a two-dimensional vector, ${\bf b}$, with projections
$b_x,b_y$. Then, for arbitrary $x$, we have
\begin{eqnarray}
\label{2}
\frac{1}{\sqrt{x^2+\vert{\bf b}\vert^2}}&=& \nonumber \\
\int
\frac{dq_xdq_y}{2\pi}&\Biggl[&\frac{\exp\left(-\sqrt{q_x^2+q_y^2}\;x\right)}{\sqrt{q_x^2+q_y^2}}
\Biggr]
\exp(iq_xb_x+iq_yb_y).\nonumber\\
\end{eqnarray}

To use this identity in Eq.~(\ref{array}), we set $b_x=nd$,
$b_y=md$. Then the summation over $n$ and $m$ can be readily
performed, yielding the sum of $\delta$-functions, i.e.
\begin{equation}
\label{3}
\left(\frac{2\pi}{d}\right)^2\sum_{p,l}\delta\Biggl(q_x-\frac{2\pi
p}{d}\Biggr) \delta\Biggl(q_y-\frac{2\pi l}{d}\Biggr),
\end{equation}
where $p$ and $l$ assume all integer values. After that the rhs in
Eq.~(\ref{EQN1}) acquires the form
\begin{equation}
\label{4} \int_0^{h}dz^{\prime}\rho(z^{\prime}){\cal
S}(z,z^{\prime}),
\end{equation}
where $S(z,z^{\prime})$ is given by
\begin{eqnarray}
\label{5} {\cal
S}(z,z^{\prime})=\frac{2\pi}{d^2}\Bigl(z+z^{\prime} -\vert
z-z^{\prime}\vert \Bigr)+ \qquad \qquad \qquad \quad\qquad \nonumber \\
%\qquad \qquad \qquad \quad
%\sum_{p,l\neq 0,0}\frac{1}{d\sqrt{p^2+l^2}}\cdot\nonumber \\
\sum_{p,l\neq 0,0}\frac{1}{d\sqrt{p^2+l^2}}\left \{
\exp\Biggl[\;-\left(\frac{2\pi}{d}\right)\;\sqrt{p^2+l^2}\;\sqrt{(z-z^{\prime})^2}\;\Biggr]\right.
\nonumber \\
\left.-\exp\Biggl[\;-\left(\frac{2\pi}{d}\right)\;\sqrt{p^2+l^2}\;\sqrt{(z+z^{\prime})^2}\;\Biggr]\right
\}.\qquad
\end{eqnarray}
The first term in (\ref{5}) describes the continuous limit and
comes from $p=l=0$ in Eq.~(\ref{3}). It coincides with the kernel,
$S_{ext}(z,z^{\prime})$, defined by Eq.~(\ref{S-result}). The
remaining sum over $p$ and $l$ recovers the kernel $S_0$ in
Eq.~(\ref{basic}). The easiest way to see this is to replace the
sums over $p$ and $l$ by corresponding integrals, which would
immediately yield $S_0(z,z^{\prime})$. However, such a replacement
is justified only  when the large number of terms contribute to
the sum. This is the case only when the condition $\vert
z-z^{\prime}\vert \lesssim d$ is met. For $\vert z-z^{\prime}\vert
\gtrsim d$ the sum over $p$ and $l$ in Eq.~(\ref{5}) is dominated
by the terms $p=0$,~$l=\pm 1$ and $l=0$,~ $p=\pm 1$, and is {\em
exponentially} decaying function of $\vert z~-z^{\prime}\vert$.
This suggests that $S_0(z,z^{\prime})$ should be substituted into
Eq.~(\ref{4}), in which the integration over $z^{\prime}$ should
be restricted to the interval $\vert z-z^{\prime}\vert \lesssim
d$. Within this interval, $\rho(z^{\prime})$ in the integrand of
Eq.~(\ref{4}) can be replaced by $\rho(z)$. The remaining integral
yields $2{\cal L}_d=2\ln(d/r)$, similarly to
Eq.~(\ref{approximate}) with $h$ replaced by $d$. The product
$2{\cal L}_d\rho(z)$ is nothing but the first term in the rhs of
Eq.~(\ref{EQN1}).

The restriction of the integration interval in Eq.~(\ref{4}) to
$\vert z-z^{\prime}\vert \lesssim d$ for the part of
$S(z,z^{\prime})$, coming from the second term in Eq.~(\ref{5}),
is, in fact, a delicate step. Although this part decays as
$\exp\Bigl\{-2\pi\vert z-z^{\prime}\vert/d\Bigr\}$ outside this
interval, the behavior of $\rho(z^{\prime})$ outside this interval
is also exponential, namely, it increases as $\exp(z^{\prime}/a)$.
Therefore, the restriction of the integration interval in
(\ref{4}) is allowed only when the exponent in $\rho(z^{\prime})$
is slower, i.e. $a$ is $\gtrsim d$. However, we know from
Eq.~(\ref{depth})  that this is indeed the case.

\section{Fluctuations of induced charge density in a random array}

The conclusion drawn in Sect.~IV that the charge distribution
Eq.~(\ref{final1}) applies not only to regular but also to a
random NT array was based on the relation $a>d$ between the
penetration depth and the lattice constant. Thus, this conclusion
pertains only to the ``body'' $z \sim a$ of the distribution.
Since $\rho(z)$ falls off exponentially away from the NT tip, it
might be expected that in the tail $(h-z) \gg a$ the randomness in
NT positions would terminate the applicability of
Eq.~(\ref{final1}). To verify this fact, one can incorporate the
positional disorder into  Eq.~(\ref{EQN1}) {\em perturbatively},
i.e. to find the correction to the average $\rho(z)$ linear in the
fluctuation of the NT density.  Then the region of applicability
of Eq.~(\ref{final1}) to the random array can be established from
the condition that the {\em typical}  disorder-induced correction
is smaller than the average. This program is carried out below.

In a random array, fluctuations in the areal concentration of NTs,
$\delta {\cal N}({\bf R})= {\cal N}({\bf R})-\langle {\cal N}({\bf
R})\rangle$, lead to the fluctuations in the distribution of the
induced charge density $\delta \rho(z,{\bf R})$. We linearize
Eq.~(\ref{EQN1})
%\begin{eqnarray}
%\label{Green} \delta \rho(z,{\bf R})= \int d{\bf q}\int
%dz^{\prime} {\mbox \Huge G} (q,z^{\prime}) \exp(i{\bf q}\cdot{\bf
%R})\delta n({\bf q})
%\end{eqnarray}
and in the first order over the fluctuations obtain
\begin{equation}
\label{h} \hat{\huge {\cal H}}\!~ \bigl\{\delta\rho(z,{\bf
R})\bigr\} ={\cal F}(z,{\bf R}),
\end{equation}
where the integral operator, $\hat{\huge {\cal H}}$, is defined as
\begin{eqnarray}
\label{definition} \hat{\huge {\cal H}}\!~ \bigl\{f(z,{\bf
R})\bigr\}=4\pi{\cal
N}_0\;a^2\Bigl[f(z,{\bf R})\qquad \qquad \nonumber \\
+\frac{1}{4\pi a^2}\int \!d{\bf R^{\prime}}\int
_0^h\!dz^{\prime}{\cal S}_{ext}(z,z^{\prime};{\bf R}-{\bf
R^{\prime}})f(z^{\prime},{\bf R^{\prime}})\Bigr].\qquad
\end{eqnarray}
The rhs in Eq.~(\ref{h}) is the potential created by the
fluctuation, $\delta{\cal N}({\bf R})$, of the density of NTs with
unperturbed  charge distribution, Eq.~(\ref{final1}). As follows
from Eq.~(\ref{EQN1}),  this potential is given by
\begin{eqnarray}
\label{calF} {\cal F}(z,{\bf R})=-\int \!d{\bf R^{\prime}}~\delta
{\cal N}({\bf R^{\prime}})\int _0^h\!dz^{\prime}{\cal
S}(z,z^{\prime};{\bf R}-{\bf
R^{\prime}})\rho(z^{\prime}). \nonumber\\
\end{eqnarray}
The fact that the kernel of the integral operator, $\hat{\huge
{\cal H}}$, depends on the difference $\bigl({\bf R}-{\bf
R}^{\prime}\bigr)$ suggests transformation from $\delta{\cal
N}({\bf R})$ and $\delta \rho({\bf R},z)$
 to the Fourier harmonics $\delta\tilde{\cal N}({\bf q})$
and $\delta \rho(z,{\bf q})$, where ${\bf q}$ is the in-plane wave
vector. Upon the Fourier transform, Eq.~(\ref{h}) assumes the form
\begin{eqnarray}
\label{q-linear}
%\Biggl[\frac{1+2g{\cal L}_c}{g}\Biggr]
4\pi{\cal N}_0a^2\; \delta \rho(z,{\bf q})+\frac{2\pi {\cal
N}_0}{q}\int
_0^h\!dz^{\prime}\delta \rho(z^{\prime},{\bf q}) \qquad\quad\nonumber \\
\times\Bigl\{\exp \bigl[-\vert z-z^{\prime}\vert q\;
\bigr]-\exp\bigl[ -(z+z^{\prime})q\;\bigr]\Bigr\}={\cal F}(z,{\bf
q}),\qquad \
\end{eqnarray}
where the rhs is proportional to $\delta \tilde{\cal N}({\bf q})$
\begin{eqnarray}
\label{calF1}
  {\cal F}(z,{\bf q}) = \frac{4 \pi a\rho _0}{q(a^2q^2-1)}
  \Biggl\{\sinh(q z)\;\exp\{-q h\}  \qquad \quad\\
 \times \Bigl[\cosh({h}/{a})
 + q a \sinh({h}/{a})\Bigr]
  -a q \sinh(z/a)\Biggl\}\;\delta \tilde {\cal N}({\bf q}).\nonumber
\end{eqnarray}
The structure of the kernel in Eq.~(\ref{q-linear}) is similar to
that in the unperturbed equation Eq.~(\ref{EQN11}). It appears
that, due to this similarity, Eq.~(\ref{q-linear}) can be solved
{\em analytically} in the same way as Eq.~(\ref{EQN11}). Namely,
upon taking the second derivative from both sides,
Eq.~(\ref{q-linear}) reduces to the following  second-order
differential equation with $z$-independent coefficients
\begin{eqnarray}
\label{ODE} \delta \rho ^{\prime \prime}(z,{\bf q})-
\gamma_q^2\delta \rho (z,{\bf q})= \frac{1}{4\pi{\cal N}_0
a^2}\Big[{\cal F}^{\prime \prime}(z,{\bf
q})-q^2{\cal F}(z,{\bf q})\Bigr],\nonumber\\
\end{eqnarray}
where $\gamma_q$ is defined as
\begin{equation}
\label{gama} \gamma_q^2= q^2+\frac{1}{a^2}.
\end{equation}
Note that the rhs of  Eq.~(\ref{ODE}) can be cast in the following
simple form
\begin{eqnarray} \label{FF} {\cal F}^{\prime \prime}(z,{\bf q})-q^2{\cal
F}(z,{\bf q}) =4\pi\delta {\tilde {\cal N}}({\bf q})\rho_0
\sinh(z/a).
\end{eqnarray}
It can be now seen from Eq.~(\ref{FF}) that
$\lambda_q\delta\tilde{\cal N}({\bf q})\sinh(z/a)$ is a particular
solution of the differential equation (\ref{ODE}). However, to
find the solution of the original integral equation
(\ref{q-linear}), one has to complement the particular solution
with the solution of the homogeneous equation, i.e. to write
\begin{eqnarray}
\label{sol} \delta \rho (z,{\bf q})&=& \Bigl[\chi _q \sinh
(\gamma_q z) + \lambda _q \sinh (z/a)\Bigr] \delta{\cal N}({\bf
q})\nonumber\\&=&P(z,{\bf q}) \delta\tilde{\cal N}({\bf q}),
\end{eqnarray} and find the constants $\chi_q$ and $\lambda_q$ by
substituting Eq.~(\ref{sol}) into Eq.~(\ref{q-linear}). This
yields

%\begin{eqnarray}
%\label{gensol}
% \nonumber to remove numbering (before each equation)
% \delta \rho (z,{\bf q})&=&
%P(z,{\bf q})\delta {\tilde {\cal N}}({\bf q})\\
%&=&\Big[\chi \sinh{(\gamma_q z)}- \frac{1}{{\cal N}_0a^2
%q\sinh({z}/{a})\Big]\delta \tilde {\cal N}({\bf q}),\nonumber
%\end{eqnarray}
\begin{eqnarray}
\label{set} \chi_q =\frac{2\rho_0}{{\cal
N}_0a^3q^2}\;\Biggl[\frac{\cosh({h}/{a}) +q
a\;\sinh({h}/{a})}{\gamma_q\cosh(\gamma_q h)+q \sinh(\gamma_q
h)}\Biggr],
\end{eqnarray}
\begin{equation}
\label{lambda} \lambda _q=-\frac{\rho_0}{{\cal N}_0a^2 q^2}.
\end{equation}
Note that $\lambda_q$ diverges at small $q$. However, the full
solution Eq.~(\ref{sol}) remains finite in the limit $q
\rightarrow 0$. It also satisfies the obvious condition $\delta
\rho (0,{\bf q}) =0$.

Equations (\ref{sol})-(\ref{lambda}) allow one to quantify the
effect of disorder in the NT positions on the distribution of
induced charge. The most interesting case is $h \gg a$, when this
distribution is determined by collective screening involving many
NTs. In this limit Eq.~(\ref{sol}) can be simplified by replacing
$\sinh(h/a)$ and $\cosh(h/a)$ by $\exp(h/a)$ and introducing
$z_1=(h-z)\ll h$. Then $h$ drops out from  $z_1$-dependent part of
$P({\bf q})$ in Eq.~(\ref{sol}), and we obtain
\begin{eqnarray}
\label{avrg-rho1} P(z_1,{\bf q})=\frac{\rho_0 e^{h/a}}{{\cal
N}_0a^2}\qquad\qquad\qquad \qquad\qquad\qquad\\
\times\Bigg[ \frac{\exp (-\gamma_q z_1)-\exp (-z_1/a)}{q^2}
-\frac{\exp (-\gamma_q z_1)}{(q+\gamma_q)(\gamma_q+1/a)}\nonumber
\Bigg].
\end{eqnarray}
The form (\ref{avrg-rho1}) is very convenient to study the effect
of disorder in the ``tail'', i.e. at large $z_1$. Indeed, assuming
Gaussian fluctuations in $\delta {\cal N}({\bf R})$, so that
\begin{eqnarray}
\label{noise} \langle\delta \tilde {\cal N}({\bf q_1})\delta
\tilde {\cal N}({\bf q_2})\rangle=2 \pi {\cal N}_0\delta({\bf
q_1}-{\bf q_2}),
\end{eqnarray}
the variance of random fluctuations in the induced charge density
can be expressed as follows
\begin{eqnarray}
\label{average} \langle \delta\rho(z_1)^2 \rangle &=&
\frac{1}{A}\int d{\bf R}\; \langle\delta\rho(z_1,{\bf R})^2\rangle \nonumber\\
&=& \frac{{\cal N}_0}{2 \pi} \int d{\bf q}\;P(z_1,{\bf q})^2.
\end{eqnarray}
Here $A$ is the normalization area. It is now seen  that the
$q$-dependence of $P(z_1,q)$ is dominated by the first term in
Eq.~(\ref{avrg-rho1}). The reason for this is the following. As
was explained in the beginning of this Section, the applicability
of Eq.~(\ref{final1}), obtained for the regular array, is expected
to be terminated in the random array at ``depths'' $z_1$ that are
$\geq a$. At these depths the average field is strongly
suppressed. On the other hand, for $z_1 > a$ one can use the
expansion $\gamma_q=\sqrt{q^2+1/a^2}\simeq \frac{1}{a}+\frac{a
q^2}{2}$. This, in turn, suggests that characteristic values of
the wave vector, $q$, are $q \lesssim 1/(az_1)^{1/2}\le 1/a$. Then
the typical ratio of the second and the first terms in
(\ref{avrg-rho1}) is  $q^2a^2 \ll 1$. It also follows from the
expansion of $\gamma_q$ that the main exponents in
$\delta\rho(z_1)$ and in the average $\rho(z_1)$ are the same.
Upon neglecting the second term in Eq.~(\ref{avrg-rho1}), the
$q$-dependence of $P(z_1,q)$ acquires the form $P(z_1,q)~\propto
\exp(-aq^2z_1/2)/q^2$. Then the $q$-integration in
Eq.~(\ref{average}) can be easily performed. We will present the
final result as the ratio of variance,
$\langle[\delta\rho(h-z_1)]^2\rangle$ and the square of average
charge density
\begin{eqnarray}
\label{avrg-rho2} \frac{ \langle[\delta
\rho(h-z_1)]^2\rangle}{[\rho(h-z_1)]^2}=\frac{\ln 2}{2}\;
\Biggl(\frac{{z_1 }}{{\cal N}_0\;a^3}\Biggr).
\end{eqnarray}
The above result offers the quantitative answer to the question
about fluctuations of the induced charge density due to the
randomness in the NT positions. In particular, it can be concluded
from Eq.~(\ref{avrg-rho2}) that the disorder-induced fluctuations
in the charge density are negligible, if $z_1 \lesssim {\cal
N}_0a^3$. Since this value is much bigger than $a$,
Eq.~(\ref{avrg-rho2}) confirms our earlier claim that
Eq.~(\ref{final1}) applies not only for regular, but also for the
random array. However, this applicability is limited by the
distance $z_1 \lesssim {\cal N}_0a^3$. For larger $z_1$ the
variance exceeds the average suggesting that the charge density
strongly fluctuates within the plane $z_1=const$. Note, however,
that these fluctuations are smooth with characteristic scale
$(z_1a)^{1/2}$, which is much smaller than $z_1$, but much bigger
than the penetration depth, $a$.

As a final remark of this Section, we point out that the {\em
lower} is the density of the random array the {\em bigger} is the
depth, $z_1$, down to which  Eq.~(\ref{final1}) applies, as it
follows from Eq.~(\ref{avrg-rho2}). However, the {\em magnitude}
of the decay of the charge density, $\rho(h-z_1)/\rho(h)$, is
governed by the ratio $z_1/a$. For $z_1={\cal N}_0a^3$, this ratio
depends on the density of the array only weakly (logarithmically).

\section{Implications for field emission}

\subsection{Single NT}
It is commonly accepted that the field emission current, $J$, from
the NT tip is described by the Fowler-Nordheim law\cite{fowler}
\begin{eqnarray}
\label{fowler}
\mbox{\Large$|$}\ln(J(F)/J_0)\mbox{\Large$|$}=\frac{4\sqrt{2mW^3}}{3e\hbar
\;\beta F},
\end{eqnarray}
where $J_0$ is the prefactor, $m$ is the electron mass, $W$ is the
work function, which, in principle, is dependent on the tip
geometry\cite{tip1,tip2,tip3}.
 Parameter $\beta$ is the field enhancement factor.
Various applications of the field emission from NTs are based on
the fact that $\beta$ is large as a result of the NT geometry,
more specifically, due to the large ratio $h/r$. The expression
for the enhancement factor routinely used in the fitting the
experimental $I$-$V$ curves\cite{latest} is $\beta =Ch/r$, where
$C\sim 1$ depends on specific geometry of the tip. Within the
Thomas-Fermi description of the induced charge distribution,
outlined in Sect. II, the expression for field at a distance,
$z_1$, from the NT tip is given by the derivative of the
potential, $\phi(z_1)$, created by the induced charges
\begin{eqnarray}
\label{enhancement} F_{ind}(z_1)= \frac
{d\phi(z_1)}{dz_1}=\frac{d}{dz_1}\int_0^h dz\; \rho(z) S_0(z,z_1),
\end{eqnarray}
where $\rho(z)$ is given by Eq.~(\ref{single}). Then the
evaluation of the integral (\ref{enhancement}) yields
\begin{eqnarray}
\label{enhancement1} \frac{F_{ind}(z_1)}{F}=\Biggl(\frac{h}{2{\cal
L}_h r}\Biggr)\min \bigl\{1, r/z_1\bigr\},
\end{eqnarray}
where we had assumed $F_{ind}\gg F$. It is seen from
Eq.~(\ref{enhancement1}) that the enhancement factor indeed has
the conventional form, $\beta =Ch/r$, with $C\approx (2{\cal
L}_h)^{-1}$ for $z_1 \lesssim r$, but it falls off with increasing
$z_1$. This suggests that for low enough applied fields, when the
electron tunnelling length $\sim W/F_{ind}$  exceeds  $r$, the
$I$-$V$ characteristics deviates from the  Fowler-Nordheim law. In
order to estimate this deviation, we substitute
\begin{eqnarray}
\label{phi} \phi(z_1)=\frac{Fh}{2{\cal
L}_h}\Biggl\{\frac{z_1}{r}\Theta(r-z_1)+\Bigl[1-\ln(r/z_1)\Bigr]
\Theta(z_1-r)\Biggr\}\nonumber \\
\end{eqnarray}
into the tunnelling action
\begin{eqnarray}
\label{current} \mbox{\Large$|$} \ln(J(F)/J_0) \mbox{\Large$|$}
 =\frac{2 \sqrt{2m}}{\hbar}
\int_0^{z_t}\!dz_1\sqrt{W-e\phi (z_1)},\nonumber\\
\end{eqnarray}
where $z_t$ is the turning point at which the expression under the
square root is zero. In (\ref{current}) we had neglected the bare
potential $eFz_1$. It is now convenient to measure the electric
field in terms of $F_0$, defined as $F_0=W/e\beta r= 2W{\cal
L}_h/eh$. The integral in Eq.~(\ref{current}) can be reduced to
the error function, $\mbox{erf}\;(x)$, after which
Eq.~(\ref{current}) acquires the form
\begin{eqnarray}
\label{emission}
\mbox{\Large$|$}\ln(J(F)/J_0)\mbox{\Large$|$}=\frac{4\sqrt{2mW^3}}{3e\hbar
\;\beta F_0}G(F_0/F),
\end{eqnarray}
where the dimensionless function $G(\tau)$ is defined as
\begin{eqnarray}
\label{G_function}
G(\tau)=\tau,\qquad \tau <1; \qquad\qquad\qquad\qquad\nonumber\\
G(\tau)=\tau-\Bigl(\tau+\frac{1}{2}\Bigr)\left(1-\frac{1}{\tau}\right)^{1/2}+\qquad\qquad
\nonumber\\
\frac{3}{4}\sqrt{\frac{\pi}{\tau}}\exp(\tau-1)\;\mbox{erf}\;(\sqrt{\tau-1}),\quad
\tau>1.
\end{eqnarray}
The plot of the function $G(\tau)$ is shown in Fig. 2. Strictly
speaking, the Fowler-Nordheim region, corresponds to $\tau < 1$,
where the slope of $G(\tau)$ is identically unity. However,
$G(\tau)$ can be linearized around $\tau >1$, where the slope is
larger. For example, at $\tau =2$ the slope is $\approx 2$. This
can be interpreted as a two-times reduction of the enhancement
factor, $\beta$, in Eq.~(\ref{emission}). A significant reduction
of the enhancement factor ({\em e.g.}, 30 times) occurs around
$\tau\approx 5$. It should be noted, that, since
Eqs.~(\ref{emission}) and (\ref{G_function}) were derived
neglecting the bare potential, their applicability is limited by
$\tau<\tau_{max}$, where $\tau_{max}$ corresponds to the applied
field $F=F_0/\tau_{max}$, for which the turning point, $z_t$, in
the tunnelling action Eq.~(\ref{current}) reaches the value $W/F$.
The latter condition can be rewritten in the form
$(\tau_{max}-1)=\ln(h\tau_{max}/r)$, yielding $\tau_{max}\approx
\ln(h/r)={\cal L}_h$. For $\tau >\tau_{max}$, i.e. for applied
fields $F> F_0/\tau_{max}$, the $I$-$V$ characteristics is given
by the Fowler-Nordheim law (\ref{fowler}) with $\beta=1$.

Overall, Fig. 2 indicates that, for low enough applied fields
there are significant deviations from the Fowler-Nordheim law in
the $I$-$V$ characteristics of an individual NT. For such fields
the linearity of the Fowler-Nordheim plots shows only within very
narrow interval of $F$. On the experimental side, there are
reports, {\em e.g.}, Ref. \onlinecite{longFowler}, where
applicability of the Fowler-Nordheim law was  demonstrated within
a rather wide
 (exceeding  $3$ times) interval of change of $F$.
In other reported measurements, see, {\em e.g.}, Ref.
\onlinecite{few0}, linearity of $\ln J$ vs. $1/F$ holds only
within a limited (less than $2$ times) interval of applied fields.
Whether or not Eq.~(\ref{emission}), derived for a single NT, is
suitable to fit experimental results depends crucially on the {\em
geometry} of the array, as we discuss below.

%Experimentally, there were reports
%\cite{ }.
%This means that the field in this experiments is $\lesssim F_0$.
%In other expermiments
%\cite{  },
%interval of applied fields. These experiments,
%likely, correspond to  $F\gtrsim F_0$.

\subsection{Array of NTs}

As it was mentioned in the Introduction, increasing the density of
NTs in the array leads to dramatic suppression of the enhancement
factor. To illustrate this point, consider first the array of low
density, when the tunnelling length is much smaller than the
distance between the neighboring NTs. Then the field created by
induced charges near the tip of a given NT can be calculated from
Eq.~(\ref{enhancement}), with $\rho(z)$ given by
Eqs.~(\ref{final1}), (\ref{final2}), This calculation yields the
generalization of the field enhancement factor
Eq.~(\ref{enhancement1}) to the case of the array of low density
\begin{eqnarray}
\label{enchancement2}
\frac{F_{ind}(z_1)}{F}=\Biggl[\frac{a}{2\;r{\cal L}_d
 }\tanh(h/a)\Biggr]\min \bigl\{1, r/z_1\bigr\}.
\end{eqnarray}
The above expression recovers the enhancement factor for a single
NT in the limit $a\rightarrow \infty$, or equivalently, ${\cal
N}_0\rightarrow 0$. For $a< h$, we conclude that the enhancement
factor for the array, compared to the single NT is suppressed as
\begin{eqnarray}
\label{suppressed} \frac{\beta({\cal N}_0)}{\beta(0)}=\frac{a{\cal
L}_h}{h{\cal L}_d}= \Biggl[\frac{{\cal L}_h^2} {2\pi{\cal
L}_d{\cal N}_0h^2}\Biggr]^{1/2} \ll 1.
\end{eqnarray}
We now turn to the high-density array. In such an array, the
tunnelling length of an emitted electron can exceed the distance,
${\cal N}_0^{-1/2}$, between the neighboring NTs. Then the form of
the tunnelling barrier is not anymore given by the potential
created by a single NT with charge distribution modified by
neighboring NTs. Instead, one has to use a general expression
\begin{eqnarray}
\label{pot} \phi_{{\tiny A}}(z_1)=\int_0^h\!dz\;\rho_0
\sinh[(h-z)/a]{\cal S}(z,z_1),
\end{eqnarray}
where ${\cal S}$ is the kernel defined by Eq.~(\ref{5}). The first
term in the kernel (\ref{5}) corresponds to the ``continuous''
limit, and yields a contribution $Fz_1$ to $\phi_{{\tiny
A}}(z_1)$. The second term in Eq.~(\ref{5})
 exhibits different behavior
for large $\left(\vert z-z_1\vert \gg {\cal N}_0^{-1/2}\right)$
and small $\left(\vert z-z_1\vert \ll {\cal N}_0^{-1/2}\right)$
distances. The expression for $\phi_{{\tiny A}}(z_1)$ that
captures both long, and short-distance behaviors has the form
\begin{eqnarray}
\label{pot_AR} \phi_{ {\tiny A}}(z_1)=Fz_1\qquad\qquad\qquad \qquad \qquad \qquad\qquad\qquad\qquad \nonumber\\
+\frac{Fd}{2\sqrt{2{\cal L}_d}}\Biggl\{ \frac{a\;\Theta
(z_1-d)}{2\pi a+d}\Bigl[\exp(-2\pi z_1/d)-1\Bigr]\quad
\nonumber\\
+ \Theta(d-z_1)\ln\left(\frac{d }{z_1}\right) \exp(-z_1/a)
\Biggr\}.\quad\quad
\end{eqnarray}
The long-distance behavior is described by the first term in the
square brackets. It represents the correction to the
``continuous'' first term, $Fz_1$, in (\ref{pot_AR}) due to
discreteness of the array. Clearly, the field enhancement due to
this term is negligible. It is the second term in the square
brackets that is responsible for the field enhancement. The
physics, captured by this term,
 is that at distance $z_1 < {\cal N}_0^{-1/2}$ the
tunnelling electron ``sees'' not only the NT, from which it was
emitted, but also neighboring NTs. This term, however, does not
contain the NT radius, $r$, which was set zero in the derivation
of Eq.~(\ref{pot_AR}). The dependence on $r$ can be reinstated in
the same way as for a single NT in Eq.~(\ref{phi}), namely
\begin{eqnarray}
\label{ret-phi} \phi_{{\tiny A}}(z_1)\approx
\frac{Fd}{2\sqrt{2{\cal L}_d}}\Biggl\{
\Theta(r-z_1)\frac{z_1}{r}+\qquad\qquad\qquad\nonumber\\
\Theta(z_1-r)\Theta(d-z_1)
\left[1-\ln\left(\frac{r}{z_1}\right)\right]
\exp\bigl(-[z_1-r]/a\bigr) \Biggr\},\quad
\end{eqnarray}
where we have retained only the part, responsible for the field
enhancement. The subsequent calculation of $I$-$V$ characteristics
using Eq.~(\ref{ret-phi}) is completely identical to the case of
an isolated NT. The result can be presented in the form similar to
Eqs.~(\ref{emission}), (\ref{G_function})
\begin{eqnarray}
\label{modified}
\mbox{\Large$|$}\ln(J(F)/J_0)\mbox{\Large$|$}=\frac{4\sqrt{2mW^3}}{3e\hbar
\;\beta({\cal N}_0) F_{{\tiny A}}}G_{{\tiny A}}(F_{{\tiny A}}/F),
\end{eqnarray}
\begin{eqnarray}
\label{GA} G_{{\tiny A}}(\tau)=\tau \Biggl[1-\Bigr(1-{1 \over
\tau}\Bigl)^{3/2}\Biggr]+\qquad\qquad\nonumber\\
{1\over \sqrt \tau}\int_1^{u_\tau}\!\!du\Biggl[(1+\ln u_\tau)
\exp\Bigl\{-{r\over a}(u_\tau-1)\Bigr\}- \nonumber \\ (1 + \ln u)
\exp\Bigl\{-{r\over a}(u-1)\Bigr\}\Biggr],
\end{eqnarray}
where $u_\tau$, which is related to the turning point, $z_\tau$,
as $u_\tau = z_\tau/r$,  satisfies the following equation
\begin{eqnarray}
\label{u_tau} \tau=(1+\ln u_\tau)\exp\Bigl\{-{r\over
a}(u_\tau-1)\Bigr\}.
\end{eqnarray}
The boundary value of the external field, $F_A$, corresponds to
the turning point $z_\tau=(F_A/F)r=\tau _A r\geq r$. The field
$F_A$ is related to the single NT boundary field, $F_0$, by
\begin{eqnarray}
\label{F_A} F_A=\frac{{\cal L}_d h}{{\cal L}_h a}F_0.
\end{eqnarray}
Therefore, the variable, $\tau_A$, defined above, is related to
the corresponding single NT variable, $\tau$, as
\begin{eqnarray}
\label{tau_A} \tau _A= \Biggl[\frac {2\pi{\cal L}_d{\cal
N}_0h^2}{{\cal L}_h^2}\Biggr]^{1/2}\tau .
\end{eqnarray}
The function $G_{{\tiny A}}(\tau_{{\tiny A}})$ is plotted in Fig.
2 together with the function $G(\tau)$. We see that, while the
parameter ${\cal N}_0h^2$ changes within a wide interval, the
function $G(\tau_A)$ remains close to $G(\tau)$. This means that
the {\em form} of the $I$-$V$ characteristics for the array is
similar to that for a single NT. The difference essentially
amounts to rescaling of the characteristic field, $F_0$,
by the factor ${\cal L}_dh/{\cal L}_ha$. In other words,
increasing the density of the array results in suppression
of the field emission without the change of the shape of the
$I$-$V$ characteristics. It should be pointed out that this
conclusion pertains to the array of randomly positioned, but
completely {\em identical} NTs. As we will see below, the
situation changes dramatically when the heights of NTs are
random.

%%%%%%%%%%%%%%%%%%%%%%%%%%%%%%%%%%%%%%%%%%%%%%%%%%%%%%%%%%%%%%%%%%%%%%%%%
%%%%%%%%%%%%%%%%%%%%%%%%%%%%%%%%%%%%%%%%%%%%%%%%%%%%%%%%%%%%%%%%%%%%%%%%
\begin{figure}[ht]
\centerline{\includegraphics[width=90mm,angle=0,clip]{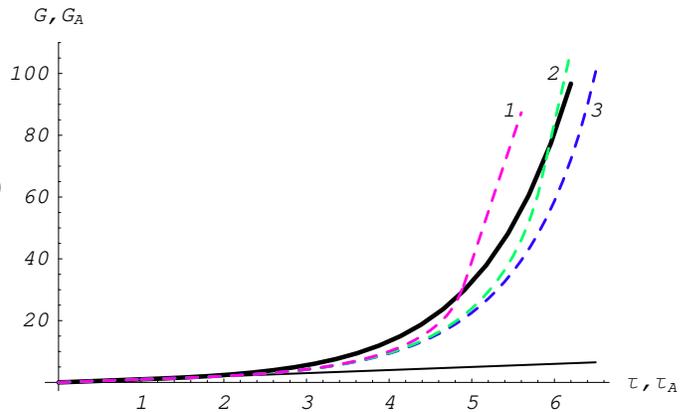}}
\caption{Thin solid line is the Fowler-Nordheim law $\vert\ln
(J/J_0)\vert \propto \tau$, where $\tau=F_0/F$. Thick solid line is
the dimensionless current-voltage characteristics $\vert\ln
(J/J_0)\vert $ vs $\tau=F_0/F$ of an individual NT plotted from
Eq.~(\ref{G_function}). Dashed curves (1), (2), and (3) are the
current-voltage characteristics, $\vert\ln J\vert$ vs
$\tau=F_{{\tiny A}}/F$
 plotted from Eq.~(\ref{GA}) for  $(h/r)=10^4$ and dimensionless array
densities ${\cal N}_0h^2= 100, 10, 1 $, respectively.}
\label{fig:2}
\end{figure}

%%%%%%%%%%%%%%%%%%%%%%%%%%%%%%%%%%%%%%%%%%%%%%%%%%%%%%%%%%%%%%%%%%%%%%%%%%
%%%%%%%%%%%%%%%%%%%%%%%%%%%%%%%%%%%%%%%%%%%%%%%%%%%%%%%%%%%%%%%%%%%%%%%%%%

\section{Discussion and Concluding remarks}
The main result of the present paper is Eq.~(\ref{final1}) that
describes the crossover of the induced charge density distribution
from a single NT to the dense regular array of NTs. We have also
demonstrated that  Eq.~(\ref{final1}) applies to the random array.
Disorder in the NT positions terminates the applicability of
Eq.~(\ref{final1}) only at large distances from the tips, where
the charge density had already dropped significantly. Concerning
the field emission, our calculations quantify a  strong
suppression of the emission current in a dense array. The field
enhancement factor falls off with the NT density, ${\cal N}_0$, as
${\cal N}_0^{-1/2}$ [see Eq.~(\ref{suppressed})]. This conclusion
might seem to contradict the majority of experiments, where high
enhancement factors for dense arrays were reported. More
precisely, in majority of experiments the dependence of emission
current on the NT density is simply not addressed, and the $I$-$V$
characteristics are interpreted basing on the properties (such as
work-function) of individual NTs. In fact, in those few papers
where this issue is addressed, the suppression of field emission
with increasing the NT density is pointed out on {\em qualitative}
level. The resolution of this contradiction, in our opinion, lies
in the fact that in realistic situations the heights of NTs in the
array are {\em widely dispersed}. To get an insight how this
dispersion in heights affects the field emission, consider  a
regular array in which one NT is higher than others by $h_1$,
which is much larger than the average NT separation, ${\cal
N}_0^{-1/2}$, but much smaller than $h$. Within the interval
$0<z<h$ the
 distribution of charge in this
``sticking out'' NT is ``enforced'' by the neighbors, and is given
by Eqs.~(\ref{final1})--(\ref{depth}). However, within the
interval $h<z<(h+z_1)$ this NT ``sees'' the rest of the array as
an equipotential plane. From this observation we immediately
conclude that, within the interval $h<z<(h+z_1)$, the charge
density in the sticking out NT is given by Eq.~(\ref{single}) with
$z$ replaced by $(z-h)$. This, in turn, suggests that the
enhancement factor of external field in the sticking out NT is
high and is equal to $h_1/2{\cal L}_{h_1}r$, as follows from
Eq.~(\ref{enhancement1}).

The above reasoning suggests that the
%It appears that the
conjecture that the field emission current is dominated by sparse
sticking out NTs
%not only
allows to account for the high values of the enhancement factor
observed in experiment. We will now demonstrate that this
conjecture also allows to explain why the dependence
$\ln(J(F)/J_0)$ follows the Fowler-Nordheim law  (\ref{fowler})
within a wide interval of $F$, while Eqs.~(\ref{emission}) and
(\ref{G_function}) predict strong deviations from
Eq.~(\ref{fowler}) as $F$ is decreased.

Obviously, the probability, $P(h_1)$, to find  within the array a
NT that sticks out by $h_1$ {\em decreases} with $h_1$. Contribution of
NTs with given $h_1$ to the emission current is determined by the
product
\begin{eqnarray}
\label{product} J_{h_1}\propto
\exp\Bigl\{\bigl[-4(2mWr^2)^{1/2}/3\hbar\bigr]
\;G(F_0/F)\Bigr\}P(h_1),\quad
\end{eqnarray}
where the first term is the tunneling action, which depends on
$h_1$ through $F_0=2W{\cal L}_{h_1}/eh_1$;  the function $G$ is
defined by Eq.~(\ref{G_function}). Since the tunneling action {\em
increases} rapidly with $h_1$, the product (\ref{product}) has a
sharp maximum at a certain optimal $h_1$. Therefore,
$\ln(J(F)/J_0)$ is determined by the logarithm of the rhs of
Eq.~(\ref{product}) taken at optimal $h_1$. The natural choice for
$P(h_1)$ is the Poisson distribution, $\exp(-h_1/H)$. We can also
use the fact that within the interval $2<\tau<10$ the function
$G(\tau)$ can be approximated  with high accuracy by the power law
\begin{equation}
\label{approximated}
 G(\tau)\approx 0.23\tau^{9/2}.
\end{equation}
Using this approximation, the optimal $h_1$ can be easily found
analytically. It is convenient to cast the final result for
$\ln(J(F)/J_0)$ in the following form
\begin{eqnarray}
\label{cast} \mbox{\Large$|$}\ln(J(F)/J_0)\mbox{\Large$|$}=
\Biggl(\frac{4\sqrt{2mW^3}}{3e\hbar \;\beta_{H} F}\Biggr)^{9/11},
\end{eqnarray}
where the ``effective enhancement factor is defined as
\begin{eqnarray}
\label{effective} \beta_H=4.8\frac{H}{r{\cal L}_H}
\Biggl(\frac{2mr^2W}{\hbar^2}\Biggr)^{7/18}.
\end{eqnarray}
Firstly, we see from Eq.~(\ref{cast}) that the $I$-$V$
characteristics is very close to the Fowler-Nordheim law, since
the exponent $9/11$ is close to $1$. This should be contrasted to
the $I$-$V$ characteristics of a single NT, for which
$\mbox{\Large$|$}\ln(J(F)/J_0)\mbox{\Large$|$} \propto F^{-9/2}$,
as follows from Eq. (\ref{approximated}).
Secondly, the effective enhancement factor (\ref{effective}) is
large and depends rather weakly on the work function $W$.
Summarizing, the dense array of NTs can exhibit the
Fowler-Nordheim field emission provided there is a sufficient
spread in the NT heights. In fact, this conclusion is in accord
with reported experimental findings. In particular, direct imaging
of emission intensity by means of scanning \cite{few0,few2} and
electron emission \cite{few1,few3} microscopy  reveals that only a
tiny portion of NTs in ($10^{-4}$ or even smaller) contributes to
the net current.

\acknowledgements

This work was supported by   NSF under Grant No. DMR-0503172
and by the Petroleum Research Fund under Grant
No.  43966-AC10.


\begin{references}

\bibitem{Science0} A. G. Rinzler, J. H. Hafner, P. Nikolaev,
L. Lou, S. G. Kim, D. Tomanek, P. Nordlander, D. T. Colbert, and
R.~ E.~ Smalley, Science {\bf 269}, 1550 (1995).

\bibitem{sourse}
P. G. Collins and A. Zettl, Appl. Phys. Lett. {\bf 69}, 1969
(1996).

\bibitem{flat0}Q. H. Wang, A. A. Setlur, J. M. Lauerhaas,
J. Y. Dai, E. W. Seelig, and R. P. H. Chang, Appl. Phys. Lett.
{\bf 72}, 2912 (1998).
%A nanotube-based field-emission flat panel display

\bibitem{flat1}K. Okano, T. Yamada, H. Ishihara, S. Koizumi,
and J. Itoh, Appl. Phys. Lett. {\bf 70}, 2201 (1997).
%Electron emission from nitrogen-doped pyramidal-shape diamond
%and its battery operation

\bibitem{flat2}J. S. Lee, K. S. Liu, and I. N. Lin,
Appl. Phys. Lett. {\bf 71}, 554 (1997).


\bibitem{flat3} Y. Chen, S. Patel, Y. Ye, D. T. Shaw,
and L. Guo Appl. Phys. Lett. {\bf 73}, 2119 (1998),
%Field emission from aligned high-density graphitic nanofibers

\bibitem{flat4}Q. H. Wang, A. A. Setlur, J. M. Lauerhaas,
J. Y. Dai, E. W. Seelig, and R. P. H. Chang, Appl. Phys. Lett.
{\bf 72}, 2912 (1998).

\bibitem{flat5}Q. H. Wang, M. Yan, and R. P. H. Chang,
Q. H. Wang, M. Yan, and R. P. H. Chang, Appl. Phys. Lett. {\bf
78}, 1294 (2001).
%Flat panel display prototype using gated carbon nanotube
%field emitters

\bibitem{flat6}H. Araki, T. Katayama, and K. Yoshino,
Appl. Phys. Lett. {\bf 79}, 2636 (2001).
%Field emission from aligned carbon nanotubes prepared by thermal
%chemical vapor deposition of Fe-phthalocyanine


\bibitem{flat7}C. J. Lee, T. J. Lee, S. C. Lyu, Y. Zhang,
H. Ruh, and H. J. Lee, Appl. Phys. Lett. {\bf 81}, 3648 (2002).
%Field emission from well-aligned
%zinc oxide nanowires grown at low temperature


\bibitem{flat8}H. Jia, Ye Zhang, X. Chen, J. Shu, X. Luo, Z. Zhang, and D.
Yu, Appl. Phys. Lett. {\bf 82}, 4146 (2003).
%Efficient field emission from single crystalline indium oxide
%pyramids







\bibitem{baughman02}
R. H. Baughman, A. A. Zakhidov, and W. A. de Heer, Science {\bf
297}. 787 (2002).
%Carbon Nanotubes--the Route Toward Applications

\bibitem{Science1}S. Fan, M. G. Chapline, N. R. Franklin,
T. W. Tombler, A. M. Cassell, and H. Dai, Science, {\bf 283}, 512
(1999).


\bibitem{latest1} see, {\em e.g.,}\\
%http://www.xintek.com/products/devices/dispaly.html
http://www.xintek.com/products/


\bibitem{latest3}
M. Chhowalla, C. Ducati, N. L. Rupesinghe, K. B. K. Teo, and G. A.
J. Amaratunga, Appl. Phys. Lett. {\bf 79}, 2079 (2001).

\bibitem{latest2} J. S. Suh, K. S. Jeong,  J. S. Lee,
and I. Han,    Appl. Phys. Lett. {\bf 80}, 2392 (2002).

\bibitem{latest4}
H. J. Lee, S. I. Moon,  J. K. Kim, Y. D. Lee,  S. Nahm, J. E. Yoo,
J. H. Han, Y. H. Lee, S. W. Hwang, and B. K. Ju, J.~Appl. Phys.
{\bf 98}, 016107 (2005).





\bibitem{theory1} Ch. Adessi and M. Devel,
Phys. Rev. B {\bf 62}, R13314 (2000).

\bibitem{theory2}S. Han, M. H. Lee, and J. Ihm,
Phys. Rev. B {\bf 65}, 085405 (2002).

\bibitem{theory3} A. Mayer, N. M. Miskovsky, and P. H. Cutler,
Phys. Rev. B {\bf 65}, 155420 (2002).

\bibitem{theory4}G. Zhou and Y. Kawazoe,
Phys. Rev. B {\bf 65}, 155422 (2002).

\bibitem{theory5}S. Han and J. Ihm,
Phys. Rev. B {\bf 66}, R241402 (2002).

\bibitem{theory6} S.-D. Liang and N. S. Xu,
Appl. Phys. Lett. {\bf 83}, 1213 (2003).

\bibitem{theory7}A. Buldum and J. P. Lu,
Phys. Rev. Lett. {\bf 91}, 236801 (2003).

\bibitem{theory8}X. Zheng, G. Chen, Z. Li, S. Deng,
and N. Xu,  Phys. Rev. Lett. {\bf 92}, 106803 (2004).

\bibitem{theory9}S.-D. Liang, N. Y. Huang, S. Z. Deng,
and N. S. Xu,  Appl. Phys. Lett. {\bf 85}, 813 (2004).

\bibitem{theory10} J. Peng, Z. Li, C. He, S. Z. Deng,
 N. S. Xu, X. Zheng, and G. Chen,
 Phys. Rev. B {\bf 72}, 235106 (2005).

\bibitem{remark}
This is the consequence of the fact that the induced charge
density depends very weakly (logarithmically) on the NT cross
section, $\pi r^2$. Then, two sufficiently close NTs can be viewed
as a single NT with cross section $2\pi r^2$.




\bibitem{Nilsson}
L. Nilsson, O. Groening, C. Emmenegger, O. Kuettel, E. Schaller,
L. Schlapbach, H. Kind, J.-M. Bonard, and K. Kern, Appl. Phys.
Lett. {\bf 76}, 2071 (2000).

\bibitem{Manohara} H. M. Manohara, M. J. Bronikowski, M. Hoenk, B. D. Hunt, and P. H.
Siegel, , J. Vac. Sci. Technol. B {\bf 23}, 157 (2005).



\bibitem{we} E.~G.~Mishchenko and M.~E.~Raikh,
preprint cond-mat/0507115.

\bibitem{chinese}Z. Li and W. Wang, preprint cond-mat/0603509.


\bibitem{rotkin} K. A. Bulashevich and S. V. Rotkin,  JETP Lett. {\bf 75}, 205 (2002).

\bibitem{blanter}  S. Sapmaz, Ya. M. Blanter, L. Gurevich, H. S. J. van der Zant,
Phys. Rev. B {\bf 67}, 235414 (2003).

\bibitem{fowler} R.~H.~Fowler and L.~W.~Nordheim,
Proc.~R.~Soc.~London, Ser. A {\bf 119}, 173 (1928).

\bibitem{tip1}G. Zhou, W. Duan, and B. Gu,
Phys. Rev. Lett. {\bf 87}, 095504 (2001).


\bibitem{tip2} Ji Luo, L.-M. Peng, Z. Q. Xue, and J. L. Wu,
Phys. Rev. B {\bf 66}, 155407 (2002).


\bibitem{tip3} S. Han and J. Ihm,
Phys. Rev. B {\bf 66}, 241402 (2002).



\bibitem{latest} see, {\em e.g.}, the most recent
papers J.~-M.~Bonard, K.~A.~Dean, B.~F.~Coll, and C.~Klinke,
Phys.~Rev.~Lett. {\bf 89}, 197602 (2002); \\
J.~Y.~Huang, K.~Kempa, S.~H.~Jo, S.~Chen, and Z.~F.~Ren,
Appl.~Phys.~Lett.  {\bf 87},
053110 (2005); \\
A.~L.~Musatov, K.~R.~Izrael'yants, A.~B.~Ormont, A.~V.~Krestinin,
N.~A.~Kiselev, V.~V.~Artemov, O.~M.~Zhigalina,
and Yu.~V.~Grigoriev, {\em ibid.} {\bf 87}, 181919 (2005);\\
J.~C.~She, S.~Z.~Deng, N.~.S.~Xu, R.~H.~Yao, and J.~Chen, {\em
ibid.} {\bf 88}, 013112 (2006);\\
N.~S.~Ramgir, I.~S.~Mulla, K.~Vijayamohanan, D.~J.~Late,
A.~B.~Bhise, M.~A.~More, and
D.~S.~Joang, {\em ibid.} {\bf 88}, 042107 (2006),\\
B. Wang, Y. H. Yang, C. X. Wang, N. S. Xu, and G. W. Yang,
J.~Appl. Phys. {\bf 98}, 124303 (2005).

\bibitem{longFowler}O. Gr\"{o}ning, O. M. K\"{u}ttel, Ch. Emmenegger, P. Gr\"{o}ning,
and L. Schlapbach, J. Vac. Sci. Technol. B {\bf 18}, 665 (2000).


\bibitem{few0}J.-M. Bonard, K. A. Dean, B. F. Coll, and C. Klinke, Phys.
Rev. Lett. {\bf 89}, 197602 (2002).


\bibitem{few2} V. I. Merkulov, D. H. Lowndes, and L. R. Baylor,
J.~Appl. Phys. {\bf 89}, 1933 (2001).

\bibitem{few1}S. Gupta, Y. Y. Wang, J. M. Garguilo, and R. J. Nemanich,
Appl. Phys. Lett. {\bf 86}, 063109  (2005).

\bibitem{few3} H. J. Lee, Y. D. Lee, W. S. Cho, B. K. Ju, Y.-H. Lee, J. H.
Han, and J. K. Kim,  Appl. Phys. Lett. {\bf 88}, 093115 (2006).

\end{references}
\end{document}